\documentclass[prd,aps,twocolumn,floatfix]{revtex4}

\usepackage{amssymb,amsmath,graphicx}
\usepackage[backref,breaklinks,colorlinks,citecolor=blue]{hyperref}
\usepackage[all]{hypcap}
\usepackage[usenames]{color}
\usepackage{natbib}
\def\jnl@style{\it}
\def\aaref@jnl#1{{\jnl@style#1}}

\def\aaref@jnl#1{{\jnl@style#1}}

\def\aj{\aaref@jnl{AJ}}                   
\def\araa{\aaref@jnl{ARA\&A}}             
\def\apj{\aaref@jnl{ApJ}}                 
\def\apjl{\aaref@jnl{ApJ}}                
\def\apjs{\aaref@jnl{ApJS}}               
\def\ao{\aaref@jnl{Appl.~Opt.}}           
\def\apss{\aaref@jnl{Ap\&SS}}             
\def\aap{\aaref@jnl{A\&A}}                
\def\aapr{\aaref@jnl{A\&A~Rev.}}          
\def\aaps{\aaref@jnl{A\&AS}}              
\def\azh{\aaref@jnl{AZh}}                 
\def\baas{\aaref@jnl{BAAS}}               
\def\jrasc{\aaref@jnl{JRASC}}             
\def\memras{\aaref@jnl{MmRAS}}            
\def\mnras{\aaref@jnl{MNRAS}}             
\def\pra{\aaref@jnl{Phys.~Rev.~A}}        
\def\prb{\aaref@jnl{Phys.~Rev.~B}}        
\def\prc{\aaref@jnl{Phys.~Rev.~C}}        
\def\prd{\aaref@jnl{Phys.~Rev.~D}}        
\def\pre{\aaref@jnl{Phys.~Rev.~E}}        
\def\prl{\aaref@jnl{Phys.~Rev.~Lett.}}    
\def\pasp{\aaref@jnl{PASP}}               
\def\pasj{\aaref@jnl{PASJ}}               
\def\qjras{\aaref@jnl{QJRAS}}             
\def\skytel{\aaref@jnl{S\&T}}             
\def\solphys{\aaref@jnl{Sol.~Phys.}}      
\def\sovast{\aaref@jnl{Soviet~Ast.}}      
\def\ssr{\aaref@jnl{Space~Sci.~Rev.}}     
\def\zap{\aaref@jnl{ZAp}}                 
\def\nat{\aaref@jnl{Nature}}              
\def\iaucirc{\aaref@jnl{IAU~Circ.}}       
\def\aplett{\aaref@jnl{Astrophys.~Lett.}} 
\def\apspr{\aaref@jnl{Astrophys.~Space~Phys.~Res.}}
\def\bain{\aaref@jnl{Bull.~Astron.~Inst.~Netherlands}} 
\def\fcp{\aaref@jnl{Fund.~Cosmic~Phys.}}  
\def\gca{\aaref@jnl{Geochim.~Cosmochim.~Acta}}   
\def\grl{\aaref@jnl{Geophys.~Res.~Lett.}} 
\def\jcp{\aaref@jnl{J.~Chem.~Phys.}}      
\def\jgr{\aaref@jnl{J.~Geophys.~Res.}}    
\def\jqsrt{\aaref@jnl{J.~Quant.~Spec.~Radiat.~Transf.}}
\def\memsai{\aaref@jnl{Mem.~Soc.~Astron.~Italiana}}
\def\nphysa{\aaref@jnl{Nucl.~Phys.~A}}   
\def\physrep{\aaref@jnl{Phys.~Rep.}}   
\def\physscr{\aaref@jnl{Phys.~Scr}}   
\def\planss{\aaref@jnl{Planet.~Space~Sci.}}   
\def\procspie{\aaref@jnl{Proc.~SPIE}}   

\def\be {\begin{equation}}
\def\ee {\end{equation}  }
\def\beq{\begin{eqnarray}}
\def\eeq{\end{eqnarray}  }
\def\bi {\begin{itemize} }

\def\ei {\end{itemize}   }
\def\RE {I\kern-6pt R    }
\def\Z  {Z\kern-13pt Z   }
\def\be {\begin{equation}}
\def\ee {\end{equation}  }
\def\beq{\begin{eqnarray}}
\def\eeq{\end{eqnarray}  }

\def\eeq{\end{eqnarray}  }

\def\apjl{Ap. J. Letters}
\def\mnras{MNRAS}
\def\physrep{Physics Reports}

\newcommand{\had}{{\sc had}}

\begin{document}

\title{Maxwell-Dilaton Dynamics}

\author{Steven L. Liebling}
\affiliation{
Department of Physics, Long Island University, Brookville, New York 11548, USA
}

\date{\today}

%
%
\begin{abstract}
The dynamics of Maxwell-dilaton theory in Minkowski spacetime are studied
using fully nonlinear, numerical evolutions. This model represents
the flat-space sector of Einstein-Maxwell-Dilaton theory which has
attracted interest recently because it is a well-posed alternative to general
relativity, and it also represents the abelian sector of Yang-Mills-Dilaton.
As such, understanding its dynamics may shed light on the dynamics of
the respective larger systems. In particular, we study electric, magnetic,
and dyonic monopoles as well as the flux tubes studied previously by  Gibbons and Wells.
Some scenarios produce large gradients that an increasing adaptive mesh refinement fails to resolve. This behavior is suggestive, although far from 
conclusive, that the growth leads to singularity formation.
No sharp transition between singularity formation and either dispersion or stationarity is found, unlike other nonlinear systems that have demonstrated behavior similar
to black hole critical behavior at such transitions.

\end{abstract}

\maketitle

%
%
\section{Introduction}
\label{sec:intro}

The Maxwell-Dilaton system contains an electromagnetic field
coupled non-minimally to a scalar field. The dilaton acts as
something of a scalar, attractive ``gravity'' allowing for compact
solutions such as those found by Morris~\cite{Morris:2006pj}~and Gibbons
and Wells~\cite{Gibbons:1993qy}.

This system also represents a particular sector of the more general
Einstein-Yang-Mills-Dilaton model. The model studied here
results from restricting 
to flat-space and U(1) abelian gauge field~\cite{Bizon:1992gi}.
If one instead allows for curved space, one has the Einstein-Maxwell-Dilaton
model studied recently~\cite{Hirschmann:2017psw,
Rocha:2018lmv,
Khalil:2018aaj,
Brito:2018hjh,
Pacilio:2018gom,
McCarthy:2018zze,
Julie:2017rpw,
Julie:2017ucp,
Jai-akson:2017ldo}.
Understanding the dynamics of this simpler system may help elucidate aspects
of the more general system.

This nonlinear system is also interesting in its own right. The monopole and flux
tube solutions, as found by Gibbons and Wells~\cite{Gibbons:1993qy}, can
be used as initial data to study their stability properties. The instability
found in the fully relativistic model of Ref.~\cite{Hirschmann:2017psw},
because its analysis relied on the equation of motion for the dilaton without relying on a particular form for the metric,  is expected to remain in this restricted
model.  Some nonlinear systems have demonstrated a type of critical behavior
similar to that found
in gravitational scalar collapse~\cite{Choptuik:1992jv}, such as
Refs.~\cite{Liebling:1999nn,Liebling:2000sy,Liebling:2002qp,Liebling:2004nr,Liebling:2005ap,Liebling:2012gv}, and so we look for such threshold behavior
here. Most of the behavior observed has been spherical and the hope with
this model is that such behavior at the threshold may be less symmetric
since the dynamics of vacuum Maxwell only occurs outside spherical symmetry.





%
%
\section{Equations of Motion}
\label{sec:eom}

We begin by varying the appropriate action in flatspace
\be
S = \int d^4x \left[  a_0 \, \bigl( \partial \phi \bigr)^2
                     +a_1 \, e^{-2\kappa \phi} \, F_{ab} F^{ab}
                       \right]
\ee
where $\phi$ is the dilaton, $F^{ab}$ is the Faraday tensor for the
electromagnetic field, and $a_0$, $a_1$, and $\kappa$ are coupling
constants. The resulting system of equations consists of the
evolution equations
\beq
\nabla_a \nabla^a \phi                          & = &
         - {a_1 \over a_0} \, \kappa \, e^{-2\kappa \phi} \, F_{ab} F^{ab}
\\
\nabla_a\left( e^{-2\kappa\phi} F^{ab} \right) & = & 0, 
\eeq
and the constraints
\be
\nabla_{[a} F_{bc]} = 0 .
\ee
Consistent with Ref.~\cite{Gibbons:1993qy}, we 
choose $a_0=1/2$ and $a_1 = 1/4$.
Adopting Cartesian coordinates, we define
$\Pi \equiv \frac{\partial\phi}{\partial t}$ so that
we can re-express the evolution equations in first-order differential form
as
\beq
E_{x,t} & = & B_{z,y} - B_{y,z}
            + 2\kappa \Bigl(   \Pi \, E_x
                             - \phi_{,y} \, B_z
                             + \phi_{,z} \, B_y
                      \Bigr)
\\
E_{y,t} & = & B_{x,z} - B_{z,x}
            + 2\kappa \Bigl(   \Pi \, E_y
                             + \phi_{,x} \, B_z
                             - \phi_{,z} \, B_x
                      \Bigr)
\\
E_{z,t} & = & B_{y,x} - B_{x,y}
            + 2\kappa \Bigl(   \Pi \, E_z
                             - \phi_{,x} \, B_y
                             + \phi_{,y} \, B_x
                      \Bigr)
\\
B_{x,t} & = & E_{y,z} - E_{z,y}
\\
B_{y,t} & = & E_{z,x} - E_{x,z}
\\
B_{z,t} & = & E_{x,y} - E_{y,x}
\\
\phi_{,t} & = & \Pi \\
\Pi_{,t}  & = &\phi_{,xx}
             + \phi_{,yy}
             + \phi_{,zz}
             + \kappa e^{-2\kappa\phi}   \times \nonumber \\*
       & &    \left[     B_x{}^2 + B_y{}^2 + B_z{}^2
                       - E_x{}^2 - E_y{}^2 - E_z{}^2   \right] .
\eeq
Commas within a subscript indicate partial derivatives with respect to
the subsequent coordinate so that $\phi_{,y}$ is equivalent to
$\partial \phi / \partial y$.

We can compute the energy as an integral over space of the energy density $\rho$ 
\beq
\rho    & = & \rho_\phi + \rho_E + \rho_B
\label{eq:rho}
\eeq
where the contributions have been separated individually as
\beq
\rho_\phi    & = & {1\over 2} \left( \Pi^2 + \phi_{,x}^2 + \phi_{,y}^2 + \phi_{,z}^2 \right) \nonumber \\
\rho_E       & = & {e^{-2 \kappa \phi} \over 2} \left( 
                    E_x^2 + E_y^2 + E_z^2 \right) \nonumber \\
\rho_B       & = & {e^{-2 \kappa \phi} \over 2} \left( 
                      B_x^2 + B_y^2 + B_z^2 \right).
\label{eq:rho2}
\eeq
The dilaton differentiates this model from electrovacuum and provides for
the conserved charge
\be
Q = {\kappa \over 2\pi} \int dx \, dy \, dz \,
       \Bigl[   \phi_{,x} E_x
              + \phi_{,y} E_y
              + \phi_{,z} E_z
       \Bigr].
\label{eq:q}
\ee
Initial data must be solutions of the two constraint equations
\beq
0 & =  &  \bigl( e^{-2\kappa\phi} E_{x} \bigr)_{,x}
        + \bigl( e^{-2\kappa\phi} E_{y} \bigr)_{,y}
   + \bigl( e^{-2\kappa\phi} E_{z} \bigr)_{,z}  \\
0 & = & B_{x,x} + B_{y,y} + B_{z,z}.
\eeq
The evolution equations preserve the constraints in the sense that the
solution at a given time generated from integrating the evolution
equations will also solve the constraint equations. However, numerically
any deviations from the constraints could in principle grow. This is
called a {\em free evolution} and contrasts with a {\em constrained 
evolution} in which the constraints equations are used in place of an
equal number of evolution equations. As such, we can monitor the
constraint {\em residual} which is an absolute, but generally arbitrary,
measure of the extent to which the solution at a given time fails to
solve the constraint equations. Numerical data presented below suggests
that residuals do not grow significantly for the time scales considered
here. It should be noted that methods from the field of computational 
magnetohydrodynamics such as a {\em divergence cleaning} or {\em constrained
transport} could be used to control such growth.

We study the evolution of different initial data and describe that data
in the following sections describing the results.
The constant $\kappa$ is equivalent to the $\alpha_0$ of Ref.~\cite{Hirschmann:2017psw}, and the results below adopt $\kappa=1$ (see Ref.~\cite{Gibbons:1993qy} for a 
discussion explaining why a change to its numerical value has no physical significance).

\begin{figure}
\includegraphics[width=3.5in]{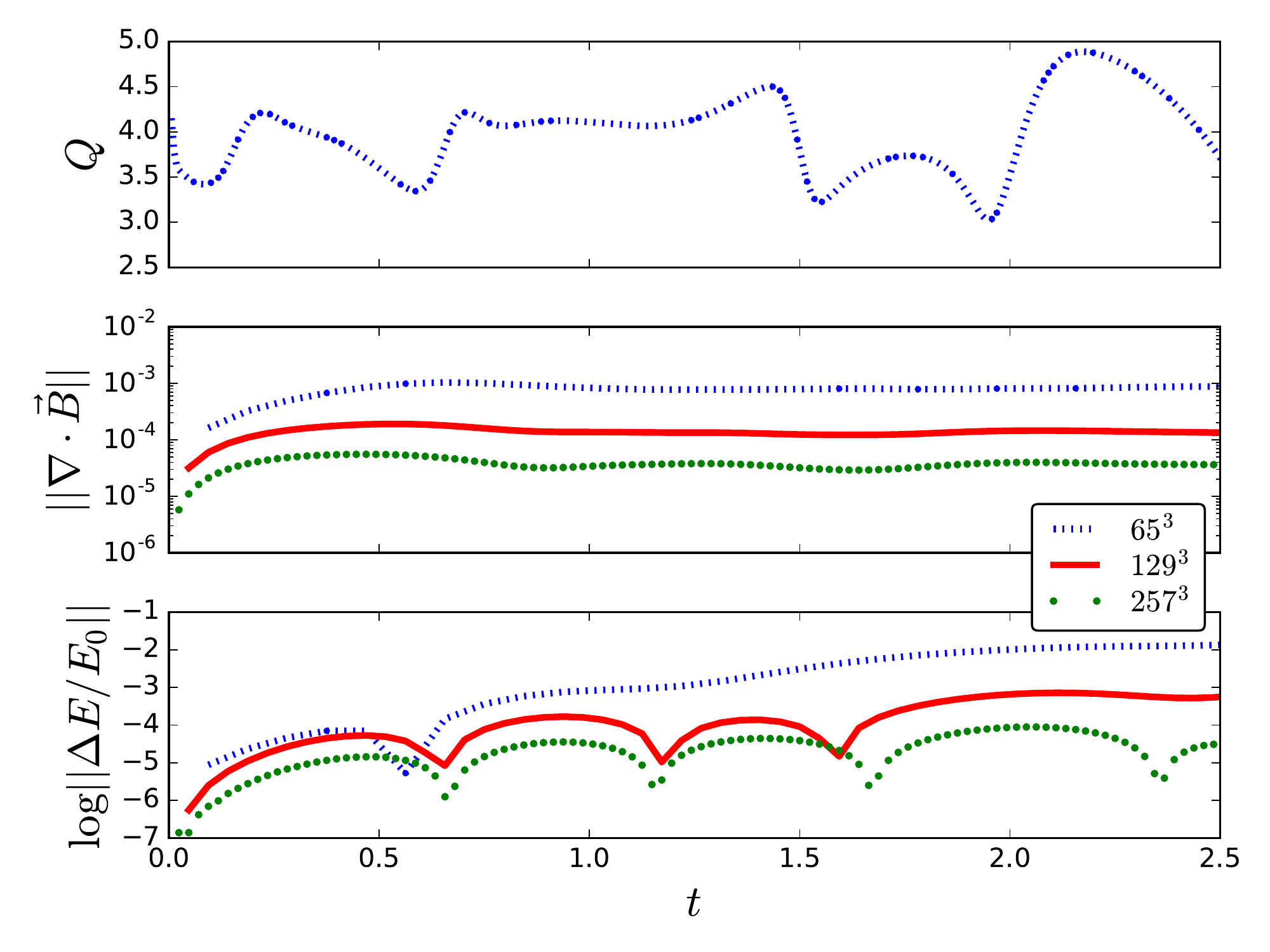}
\caption{Demonstration of convergence for the Gaussian initial data in Eqs.~\eqref{eq:gaussian1}-\eqref{eq:gaussian2} with
5 levels of FMR refinement. (Top:) The convergence order is computed over just
the finest level and demonstrates convergence of the $x$-component of $\vec{B}$.
(Middle:) The norm of the divergence of the magnetic field decreases with
resolution. 
(Bottom:) The fractional energy change similarly decreases with resolution.
All these indications suggest that the code is convergent and consistent.
}\label{fig:convergence}
\end{figure}
%
%
\section{Implementation}
\label{sec:imp}

We solve these equations using the distributive, adaptive mesh infrastructure
for finite differences \had~\cite{had_webpage}.
We use fourth-order accurate, center differences in a method of lines scheme with a third-order accurate Runge-Kutta time integrator.
We use Kreiss-Oliger-like dissipation but with high order derivatives as a mild low-pass filter to mitigate noise, as is fairly common in numerical relativity codes.

We present the results of a particular numerical test in Fig.~\ref{fig:convergence}. For these tests, we evolve what we call Gaussian initial data 
which is smooth
everywhere and satisfies the constraints. Here, we set $\phi=0=\vec{B}$ and
\begin{eqnarray}
E_x & = &  0\label{eq:gaussian1}\\
E_y & = & 
            \left( \frac{ z}{\delta_z^2} \right)
            A 
               e^{-x^2/\delta_x^2}
               e^{-y^2/\delta_y^2}
               e^{-z^2/\delta_z^2}
\label{eq:gaussian0}\\
E_z & = & 
            - 
            \left( \frac{ y}{\delta_y^2 } \right)
            A 
               e^{-x^2/\delta_x^2}
               e^{-y^2/\delta_y^2}
               e^{-z^2/\delta_z^2}
\label{eq:gaussian2}
\end{eqnarray}
for real parameters $A$, $\delta_{x|y|z}$.
In the figure, we compare three different resolutions and demonstrate that
the convergence order is consistent with third order. Note that these evolutions
use fixed mesh refinement~(FMR) and the order is computed only by comparing the finest
levels. Also shown are the total divergence of the magnetic field and change
in total energy versus time. Both of these represent errors and that
the measures of error decrease with increasing resolution represents
a test of consistency for the numerical solution.

%
%
\section{Results}
\label{sec:results}

%

\subsection{Monopoles}
Another description of initial data are the monopoles of Ref.
~\cite{Gibbons:1993qy}.
The \textbf{magnetic monopole}
can be expressed in terms of a monopole charge $P$ such that
\beq
\phi & = &  \left( \frac{1}{\kappa} \right) \ln \left[
                P \kappa \left( \frac{1}{r} + \frac{1}{P \kappa} \right)
                \right] \\
%
B_r & = &  \frac{P \kappa}{r^2}.
\eeq
Because the monopole is singular at the origin, a cutoff $r_0$ is
instituted such that for $r\le r_0$, the radius used in the above
equations is instead $r_0$ so that near the origin, for example,
$B_x  =  P\kappa x/r_0^3$.

\begin{figure}
\includegraphics[width=3.5in]{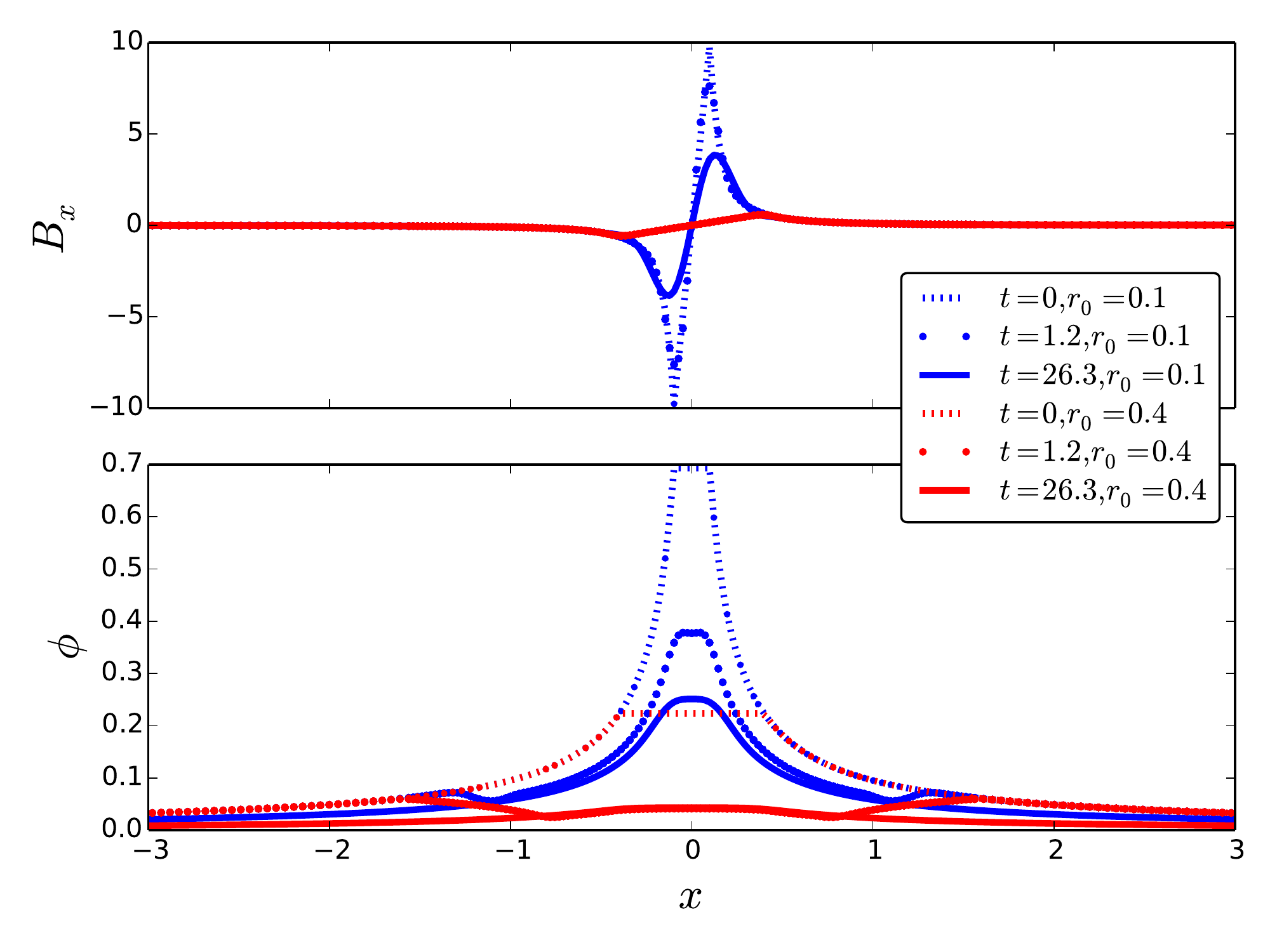}
\caption{Dynamics of the magnetic monopole. Shown are snapshots of
the $x-$component of the magnetic field and the dilaton on 
the $x$-axis for
a magnetic monopole with $P=0.1$ and radial cut-off $r_0$
equal to either $0.1$ or $0.4$. 
The magnetic field settles more quickly than the dilaton which
``sheds'' some excess field, but the overall solution settles
quickly to a stationary solution that depends on $r_0$.
}\label{fig:monosettle}
\end{figure}

The dynamics observed for magnetic monopoles is demonstrated in
Fig.~\ref{fig:monosettle}.
The magnetic monopole appears stable for all charges tried.
The effect of the cutoff appears to be that the solution ``sees''
a charge that depends on the cutoff. In particular, the dilaton
settles into different solutions dependent on $r_0$. A small
value of $r_0$ demands a higher resolution to capture the gradients,
particularly in the magnetic field (as opposed to the dilaton).


%

\begin{figure}
\includegraphics[width=3.5in]{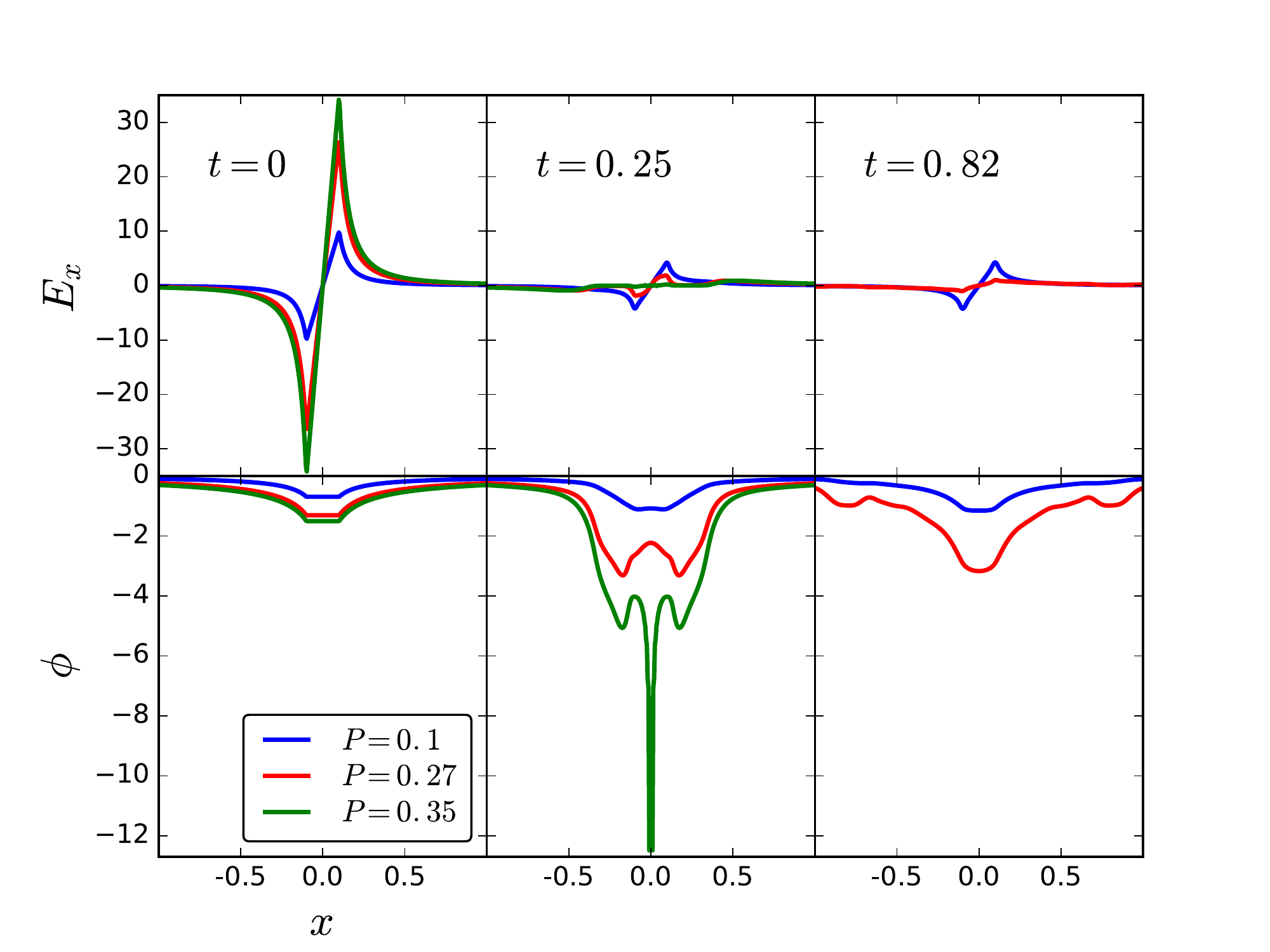}
\caption{Three regimes of dynamics for the electric monopole.
The solution at three different times is shown for three different charges.
For small charges, the monopole appears dynamically stable (see the $P=0.1$ case shown in blue).
For large charges, the dilaton grows at the center until the code no longer resolves the solution (see the $P=0.35$
case shown in green). 
Instead of a sharp transition, there appears to be a regime in which the dilaton grows but saturates or otherwise
stops growing (see the $P=0.27$ case shown in red).
Here the cut-off value, $r_0$, is equal to $0.1$ and note that the spatial
extent of the simulation extends much further than what is shown.
}\label{fig:electricmonopole}
\end{figure}

We can define an \textbf{electric monopole} similarly
\beq
\phi & = & -\left( \frac{1}{\kappa} \right) \ln \left[
                P \kappa \left( \frac{1}{r} + \frac{1}{P \kappa} \right)
                \right] \\
E_r & = &  \frac{P \kappa}{r^2}.
\eeq
Again, $P$ is the monopole charge and $r_0$ is the length scale at
which the solution is cut-off.
Three representative evolutions are shown in Fig.~\ref{fig:electricmonopole}. For small monopole charge, the solution appears stable, similar to the magnetic case. For large monopole charge, however, the solution appears unstable with the dilaton becoming more and more negative in time. 

One may expect critical
behavior to appear in between these two regimes as has been observed
in gravitational scalar collapse~\cite{Choptuik:1992jv} and in certain
non-gravitating nonlinear theories~\cite{Liebling:1999nn,Liebling:2000sy,Liebling:2002qp,Liebling:2004nr,Liebling:2005ap,Liebling:2012gv}.  However, instead there appears
to be a third regime intermediate between small and large charge in which
the dilaton becomes more negative at early times and then saturates.

Likewise, we define a \textbf{dyonic monopole} as
\beq
\phi & = & 0\\
B_r & = &  \frac{P \kappa}{r^2} \\
E_r & = &  \frac{P \kappa}{r^2}.
\eeq
Here $P$ is both the electric and magnetic monopole charge and $r_0$ is the length scale at
which the solution is cut-off.
Because we consider here only dyonic monopoles with equal electric and magnetic charges, the
natural choice is for the dilaton to vanish.

\begin{figure}
\includegraphics[width=3.5in]{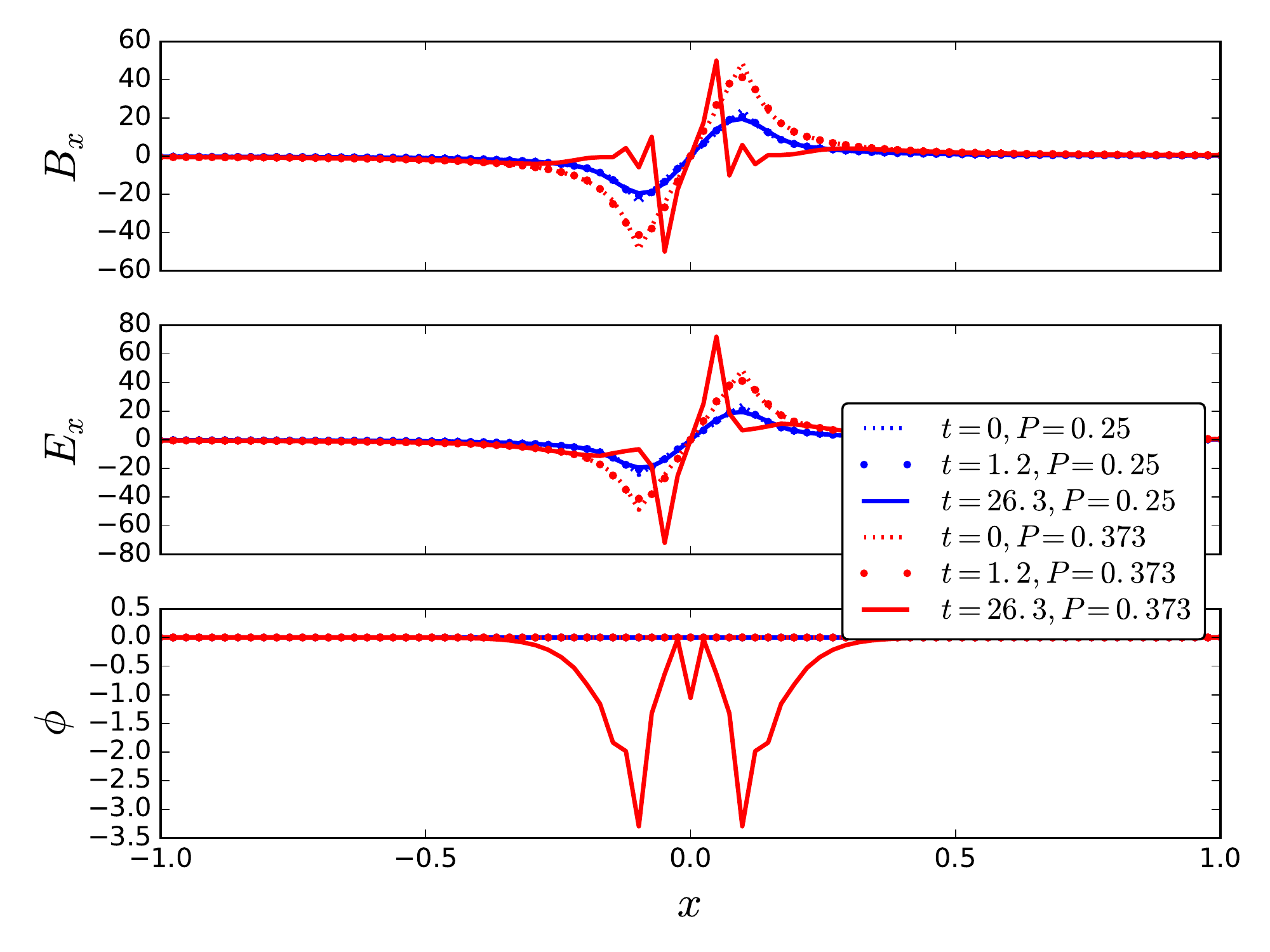}
\caption{Demonstration of instability in the dyonic monopole.
The solution at a few different times is shown for two charges,
$P=0.25$ (stable) and $P=0.373$ (unstable).
}\label{fig:unstdyon}
\end{figure}

The dynamics of these monopoles is represented in Fig.~\ref{fig:unstdyon}. Two regimes
are observed depending on the charge $P$. For small charge, the solution appears stable
whereas for large charge, the evolution indicates instability. In particular, the
dilaton becomes more and more negative while the electric
field grows larger than the magnetic field. Both these behaviors suggest that the
instability found for electric monopoles dominates the dynamics in this regime.

\subsection{Flux Tubes}

Another type of solution, namely the fluxtube as discussed in
Refs.~\cite{Gibbons:1993qy,Morris:2006pj}, describes a ``string'' of confined
magnetic flux. Here, we consider the dynamics and stability of such tubes with
generalized initial data given by
\beq
f(z) & = & A e^{-z^2/\delta^2}\\
\bar \rho & = & \sqrt{  \left( x+ f \cos kz\right)^2 
                      + \left( y+ f \sin kz\right)^2 } \\
\phi & = & \frac{1}{2\kappa} \log \left[ \frac{4a^2}{\kappa^2 H^2 \left( 1+a^2 \bar \rho^2 \right)^2} \right] \\
B_z & = & H e^{2\kappa \phi}.
\eeq
Here, the real function $f(z)$ serves to introduce a wiggle with wavenumber $k$, amplitude $A$, and width $\delta$ to the
original flux tube described by
real constants $a$ and $H$ where $H$ describes the magnetic strength of the tube.
For parameter $A=0$, one has a vertically oriented flux tube.

\begin{figure}[h]
\centerline{\includegraphics[width=3in]{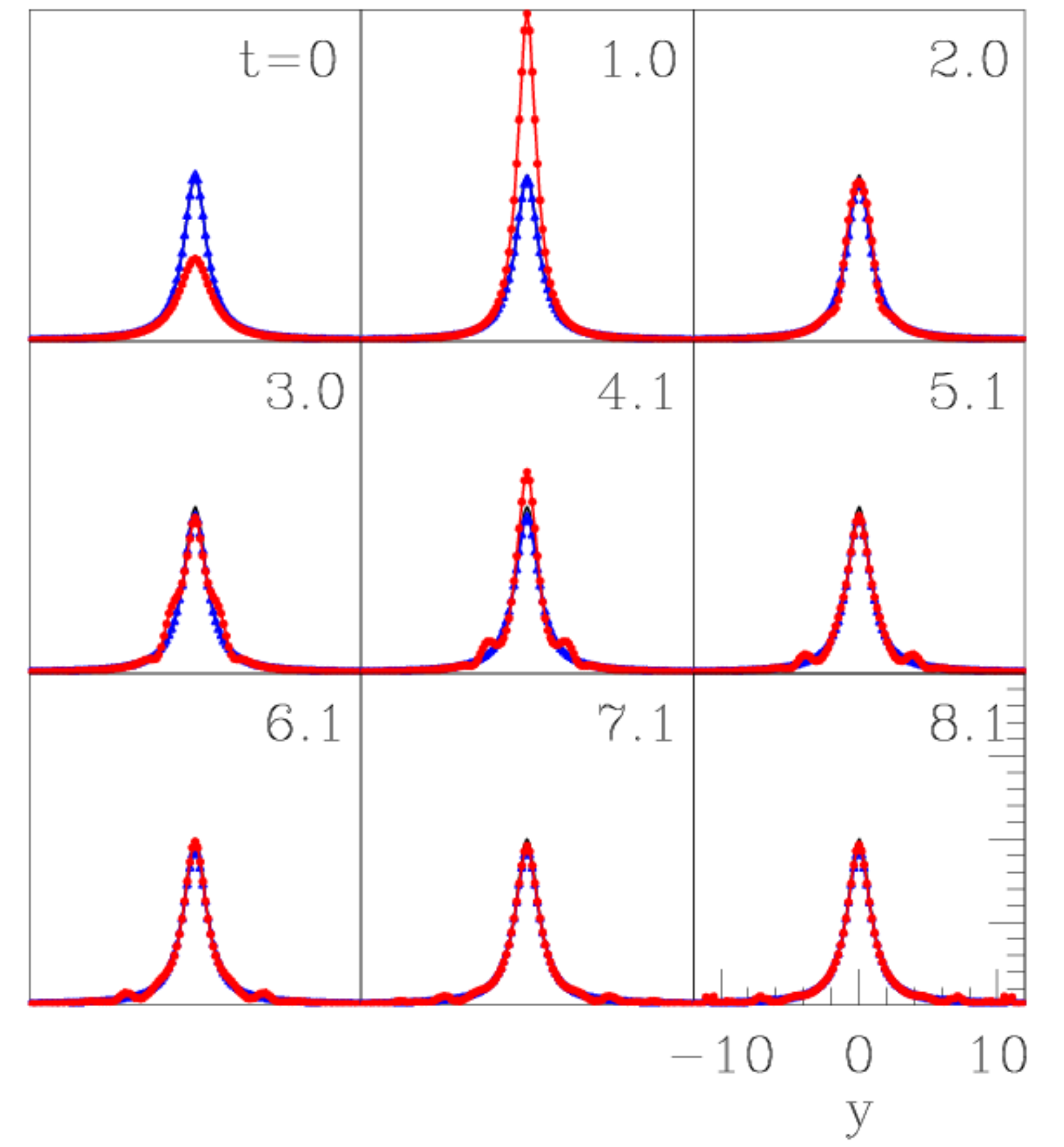}}
\caption{Evolutions of a perturbed (blue circles) and an unperturbed (red triangles) flux tube. Shown are 1D slices of the energy density $\rho$ at the times indicated for $x=0=z$. Also shown is the known, static solution (black solid line) although it is hardly visible because the other solutions overlay it. The unperturbed solution remains quite close to the static solution for the duration of
the evolution. The wiggly flux tube oscillates around the static solution with the oscillation damped quickly by outgoing radiation visible in the frames $t=3.0$ and $t=4.1$. Noisy boundary effects are becoming apparent in the last frame.
Here $H=1$ and $a=1$ for both solutions, 
and the perturbation is described by parameters: $A=1$, $k=2\pi/5$, and $\delta=4$.
}
\label{fig:wiggle}
\end{figure}

The dynamics of certain solutions are shown in Fig.~\ref{fig:wiggle}. An unperturbed,
straight tube appears stable. Note that the tube necessarily hits the boundaries
of the computational domain, and thus stability is suspected based on short evolutions before boundary effects become significant throughout the domain.
Likewise, for non-vanishing $A$, the ``wiggled'' string also appears stable with the perturbation
quickly propagating away, leaving what appears to be the unperturbed, stationary string.


\subsection{Evidence for Singularity Formation}

Our final comment concerns whether this model permits singularity formation or whether the 
large growth observed in the previously mentioned unstable cases ultimately saturates.
Numerics likely cannot fully answer this question of global existence, but it can suggest
an answer. And so instead of considering initial data which itself needs to be regularized, such
as the introduction of a cut-off scale $r_0$ with the monopoles, we instead return to
the initial data found in Eqs.~\eqref{eq:gaussian1}-\eqref{eq:gaussian2}
 and for which convergence was demonstrated in Fig.~\ref{fig:convergence}. This initial data is
smooth everywhere.

\begin{figure}[h]
\centerline{\includegraphics[width=3in]{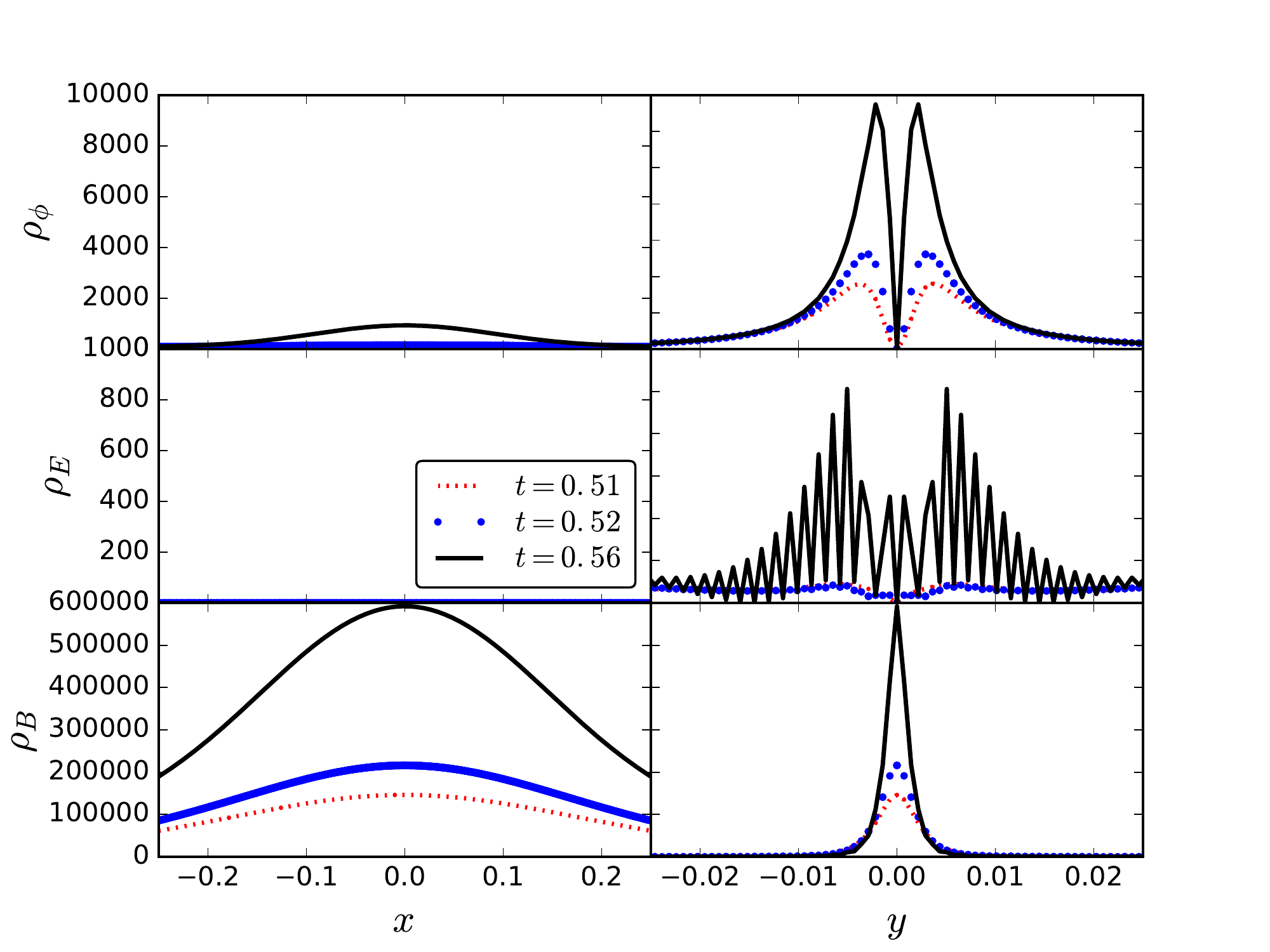}}
\caption{
Evolution  of initial data of the form found in Eqs.~\eqref{eq:gaussian1}-\eqref{eq:gaussian2} towards apparent singularity. Slices along the $x-$ and $y-$ axes of the different contributions to the energy
density for a particular evolution with Gaussian initial data.
The energy concentrates along the $x-$axis, shrinking to roughly a line with thickness of order the grid spacing. Note: (i)~the much smaller spatial bounds of the right column of plots versus
the left, and (ii)~a plot over $z-$ would be similar to that of $y-$.
Increasing the number of AMR levels resolves the concentration of energy better and delays slightly the point at which the
code cannot handle the large gradients. That increasing resolution demonstrates this same
concentration is suggestive of singularity formation.
Shown is the highest resolution case with seven levels of AMR.
}
\label{fig:singform}
\end{figure}

We characterize the dynamics in Fig.~\ref{fig:singform} for large amplitude
by showing snapshots of the various
contributions to the energy density as defined in Eq.~\eqref{eq:rho2}. The initial data
quickly evolves such that the energy concentrates along the $x$-axis. The adaptivity places
refined grids in this vicinity. However, runs with increasing number of allowed levels of adaptive mesh
refinement~(AMR) all show dynamics in which the concentration reaches the grid resolution. This
behavior suggests that the concentration occurs without limit, but of course such an
extrapolation is a guess because the continuum equations could dictate saturation at a scale
beyond the reach of these runs.

\section{Conclusion}

Numerical evolutions of various forms of initial data in the Maxwell-Dilaton system indicate various regimes of stability and instability. In particular,
an instability for certain electric dominated scenarios seen in the gravitating
case appears to carry-over to this flatspace model~\cite{Hirschmann:2017psw}. 

A particular case which demonstrated unstable growth was studied with 
increasing adaptive refinement that was unable to fully resolve the growth.
This behavior was suggestive that the growth is unbounded and will ultimately
form a singularity, although the numerics here cannot be conclusive.

Certain systems that demonstrate two disparate dynamical regimes such as 
singularity formation and stationarity or singularity versus dispersion
have also demonstrated critical behavior at the threshold similar to
black hole critical behavior~\cite{Choptuik:1992jv}. However, no
such behavior is found in this model.

%
%
\section*{Acknowledgments}
It is a pleasure to thank  David Garfinkle
for useful discussions and comments and Eric Hirschmann for suggesting
this project and for his early help. We acknowledge
support from NSF grants
PHY-0969827,
PHY-1607291,
PHY-1827573,
and
PHY-1912769
 to Long Island University.
Computations were
performed in part on  XSEDE 
computational resources.

%
%
\bibliography{./the}
\bibliographystyle{utphys}

%
%
\end{document}